\documentclass[aps,prd,showkeys,showpacs,amsfonts,amssymb,a4paper]{revtex4-2}
\usepackage[top=2.5cm,bottom=2.5cm,left=2cm,right=2cm]{geometry}
\usepackage{xcolor}
\usepackage{graphicx}
\usepackage{physics}
\usepackage[colorlinks=true,citecolor=green,linkcolor=blue]{hyperref}

\newcommand\bef{\begin{figure}[!h]}
\newcommand\eef[1]{\label{fg:#1}\end{figure}}
\newcommand\beq{\begin{equation}}
\newcommand\eeq[1]{\label{#1}\end{equation}}
\newcommand\bea{\begin{eqnarray}}
\newcommand\eea{\end{eqnarray}}
\newcommand\bet{\begin{table}}
\newcommand\eet[1]{\label{tb:#1}\end{table}}

\newcommand\fgn[1]{Fig.\,\ref{fg:#1}}
\newcommand\eqn[1]{Eq.\,(\ref{#1})}
\newcommand\scn[1]{Section \ref{sec.#1}}
\newcommand\apx[1]{Appendix \ref{sec.#1}}

\newcommand{\chc}{\langle \bar{\psi} \psi \rangle}

\newcommand{\nc}{N_c}
\newcommand{\nf}{N_f}
\newcommand{\tc}{T_c}

\newcommand{\nfm}{n_{\scriptscriptstyle F}}
\newcommand{\muq}{\mu}
\newcommand{\mpi}{m_\pi}
\newcommand{\mub}{\mu_{\scriptscriptstyle  B}}
\newcommand{\nB}{n_{\scriptscriptstyle  B}}
\newcommand{\nq}{n}
\newcommand{\tmu}{\left( T, \, \mu_{\scriptscriptstyle  B} \right)}
\newcommand{\tmq}{\left( T, \, \mu \right)}
\newcommand{\chim}{\chi_m}

\newcommand{\tuv}{\tau_{\rm \scriptscriptstyle UV}}
\newcommand{\luv}{\Lambda_{\rm \scriptscriptstyle UV}}
\newcommand{\lnjl}{\mathcal{L}^{\scriptscriptstyle \rm MF}_{\scriptscriptstyle \rm NJL}}
\newcommand{\omf}{\Omega}
\newcommand{\veff}{V_{\rm eff}}
\newcommand{\efc}{{\rm Erfc}}

\newcommand{\txt}{\textstyle}
\newcommand{\lnj}{\mathcal{L}_{\scriptscriptstyle \rm NJL}}

\begin{document}
\title{QCD Phase Diagram and the Finite Volume Fireball: A Model Study}

\author{Adiba Shaikh\footnote{Current affiliation: {\it Tata Institute of Fundamental Research, Homi Bhabha Road, Mumbai 400005, India.}}}
\email{adibashaikh9@gmail.com}
\affiliation{Department of Physics, Indian Institute of Technology Bombay, Powai, Mumbai 400076, India.}

\author{Ranjita K. Mohapatra}
\email{ranjita.iop@gmail.com}
\affiliation{Department of Physics, Rajdhani College, Bhubaneswar, Odisha 751003, India.}

\author{Saumen Datta}
\email{saumen@theory.tifr.res.in}
\affiliation{Tata Institute of Fundamental Research, Homi Bhabha Road, Mumbai 400005, India.}

\begin{abstract}
Experimental investigations of the phase diagram of strongly
interacting matter involve collisions of heavy ions at
ultrarelativistic velocities. The medium created in such a collision
is often of dimensions a few fermi, in particular in the Beam Energy
Scan experiments. An understanding of the effect of the finite volume
and the boundary is important for connecting the experimental results
to the phase diagram.

Using the Nambu Jona-Lasinio model, an effective theory for the chiral
transition of quantum chromodynamics (QCD), we have studied the effect
of the finite volume of the fireball on the transition line at finite
temperature and density using the MIT boundary condition to mimic the
condition that the system is deconfined inside. We studied the effect 
of the finite volume on the transition temperature and on number 
density and its susceptibilities. The volume
effects should be considered when looking for signatures of the phase
diagram in experiments.
\end{abstract}


\maketitle

\section{Introduction}
\label{sec.intro}
The bulk property of the strongly interacting matter has been the subject of detailed experimental investigation in recent times
\cite{ALICE:2022wpn, Arslandok:2023utm, Tannenbaum:2012ma,
  STAR:2017sal}. Strongly interacting matter is expected to manifest
itself in a deconfined, chiral symmetry-restored phase, called the
quark-gluon plasma (QGP), at very high temperatures
\cite{Aarts:2023vsf, Braun-Munzinger:2015hba}. Experimentally, this
phase has been explored in the study of the medium created in the
collision of heavy ions at ultrarelativistic velocities, in particular
in the Relativistic Heavy Ion Collider (RHIC) in BNL, USA
\cite{Tannenbaum:2012ma} and the Large Hadron Collider (LHC) in CERN,
Geneva \cite{ALICE:2022wpn}. The phase structure at high number
density is being explored in the Beam Energy Scan (BES) experiments in
RHIC \cite{STAR:2017sal} and will be further explored in future
experiments in NICA, JINR \cite{Blaschke:2017lzo} and in the CBM experiment in FAIR \cite{Almaalol:2022xwv}. The knowledge about
the phases at high density is complemented by the data from neutron
stars and their mergers \cite{Lovato:2022vgq}. On the theory side, the
properties of the deconfined state at high temperatures are
well-understood from large-scale numerical studies of quantum
chromodynamics (QCD), the theory of strongly interacting matter, using
lattice discretization \cite{Aarts:2023vsf}. Lattice studies are more
difficult at nonzero baryon number chemical potential ($\mub$);
however, the equation of state and baryon number susceptibilities have
been calculated at large temperatures and moderate values of
$\mub$. At higher values of $\mub$, or lower temperatures, one has to
rely on QCD-based models \cite{Buballa:2003qv,Fukushima:2003fw}.

The theoretical studies, in particular the lattice-discretized QCD
(LQCD) studies are for static equilibrated infinite volume
systems. On the other hand, the medium created in RHIC and LHC
experiments is rapidly evolving, and the system size is not too
large. In particular, in the BES studies the system dimension can be a
few fermi, not too large compared to the typical correlation length in
strong interaction physics. The current experimental efforts are to study QGP in smaller systems. All this makes an understanding of the volume effect important. 

In this paper, the aim is to get an idea of the effect of the finite
volume on the phase diagram for a static system; the effects of the
dynamic nature are being left for a future study. We study the phase diagram
using the Nambu Jona-Lasinio (NJL) model \cite{Buballa:2003qv}, which
is an effective model expected to capture the physics related to the
chiral symmetry, in particular, the spontaneous chiral symmetry
breaking. Since the finite temperature crossover is associated with
the melting of the chiral condensate, the NJL model is expected to
capture the basic features of the crossover.

In a study of the finite volume effects it is essential that realistic
boundary conditions are put in for the system. The actual boundary of 
the medium created in the heavy ion collision experiments is of
complicated geometry. We take the simplified MIT boundary condition
on a spherical geometry \cite{Jaffe:1989pn} (see \cite{Greiner:1995jn} 
for a pedagogic discussion). While simple, the MIT boundary condition 
captures the essential physics that inside the boundary we have a deconfined
medium. The spherical MIT bag model has been used to study the effect of the finite size of the QGP droplet on the thermal phase-space distributions of quarks and gluons \cite{Elze:1986db}.

The effect of the finite volume on the chiral transition has been
studied before \cite{Palhares:2009tf, Bhattacharyya:2012rp,
  Bhattacharyya:2014uxa, Almasi:2016zqf, Klein:2017shl, Magdy:2019frj,
  Bernhardt:2022mnx, Kovacs:2023kbv, Kovacs:2023kcn}. These studies
have used the anti-periodic (APBC) and periodic (PBC) boundary
conditions, or have put in a lower cutoff in momentum. However, these
boundary conditions do not mimic a finite-volume fireball; they are
boundary conditions used in theoretical studies to mimic an infinite-volume system within a finite-volume setup, e.g., in numerical lattice
studies. In \apx{bc} we compare the MIT boundary condition results 
with those obtained with other boundary conditions. The MIT boundary condition
has been used before, for studying the chiral crossover at $\mub=0$ on
a sphere \cite{Zhang:2019gva}, and for a rotating fermionic system in
a cylinder \cite{Chernodub:2016kxh, Chernodub:2017ref}. Here we use it
to study the phase diagram in the $\tmu$ plane and the baryon number
susceptibilities.

The plan of the paper is as follows. In \scn{methods} we outline the
basic calculations, discuss the boundary conditions, and specify the
parameters used. The change of the phase diagram with changing volume
is discussed in \scn{pdg}. The number density and the susceptibilities
in the critical region are discussed in \scn{num}. We summarize our
results in \scn{summary}. The relevant equations for the MIT boundary
condition and other details are compiled in \apx{detail}. \apx{bc} has
a discussion of other boundary conditions.

\section{Formalism and the boundary condition}
\label{sec.methods}
The NJL model is an effective model of the spontaneous chiral symmetry
breaking of QCD \cite{Klevansky:1992qe, Buballa:2003qv}. The degrees
of freedom are only quarks; the effects of the gluonic interactions are included in four-fermion interaction terms, chosen such that the 
Lagrangian is chirally symmetric.

We will work with the two-flavor, isospin symmetric theory. The NJL
Lagrangian for two degenerate flavors of light quarks is,
\begin{equation}
    \lnj={\bar \psi}(i\gamma_\mu \partial^\mu-{\hat
      m})\psi +G\left[({\bar \psi}\psi)^2
      +({\bar \psi}i\gamma_5{\bar \tau}\psi)^2\right],
      \label{NJLlag}
\end{equation}
where $\psi=\begin{pmatrix} \psi_u \\ \psi_d \end{pmatrix}$ is the
quark field, ${\hat m}=\text{diag}(m,m)$ is the degenerate light quark
mass matrix, and $G$ is the coupling strength of the effective
four-fermion interaction between quarks. In the massless limit
($m=0$), $\lnj$ has a global $SU(2)_L \times SU(2)_R \times U(1)_V$
symmetry. This symmetry is spontaneously broken to $SU(2)_V \times
U(1)_V$ as the chiral condensate $\chc$ takes a nonzero value. In the
presence of the mass term, the symmetry is also explicitly broken,
leading to a preferred vacuum in the broken phase.

We will be studying the theory in the mean-field (MF) approximation.
Writing the chiral condensate to be:
\beq
\chc \; = \; - \frac{\sigma}{2 G},
\eeq{mf}
$\lnj$ can be rewritten in the mean-field approximation as,
\beq
\lnjl \; = \; \bar{\psi} (i\gamma_\mu \partial^\mu \, - \,
M)\psi \, - \, \frac{\sigma^2}{2 G}, \qquad M \, = \, m \, 
+ \, \sigma \, .
\eeq{mfNJL}

Then the effective potential is given by,
\beq
\veff (\sigma) \; = \; - \frac{1}{V} \, \log Z \; = \; 
\frac{\sigma^2}{2 G} \, - \, \int \mathbb{D}p \, \log \, 
\det \left(\gamma \cdot p \, - \, M \right) \, ,
\eeq{veff}
and $\sigma$ can be obtained from the minimum of
$\veff (\sigma): \ \frac{\txt \partial \veff}{\txt \partial \sigma} \, = \, 0.$
Here $\mathbb{D}p$ refers to the regularized
momentum integral. 

The four-fermion term in $\lnj$ makes the theory
non-renormalizable. Hence, a suitable regularization method has to be
specified, and the results will depend on the regularization
parameter. We have used Schwinger's proper time regularization
\cite{Klevansky:1992qe, Kohyama:2015hix}, with the cutoff $\tuv =
1/\luv^2$. Therefore, our model has three parameters: $m, G$ and the
UV momentum cutoff $\luv \equiv \Lambda$. These parameters can be chosen by
specifying the pion mass, the pion decay constant and either the
chiral condensate or the chiral transition temperature ($T_\chi$). Some
suitable parameter sets, and how they affect the chiral crossover
temperature, have been discussed in Ref. \cite{Kohyama:2015hix}. The
phase diagram can be qualitatively different for the different
parameter sets. In particular, for some choices of parameters one gets
a critical point and a first-order transition at large quark number
chemical potential $\muq=\dfrac{\mub}{3}$, while for some other
choices, the transition is a crossover in the whole $\tmq$ plane. We
have chosen the following parameter set:
\begin{center} \begin{tabular}{c|c|c} 
 \hline
  $m$ (MeV) & $G$ (\text{GeV}$^{-2}$) & $\luv$ (MeV) \\ 
 \hline
 15 & 17.2 & 645 \\ 
 \hline
\end{tabular} \end{center}
This parameter set was already explored in
Ref. \cite{Kohyama:2015hix}. It has a first-order transition line at
large $\muq$, ending at a critical point at $\mu_c \approx$ 340
MeV. At $\muq = 0$, it shows a chiral transition at $T_\chi \approx$
186 MeV. These values are larger than those expected for QCD: lattice
calculations find $T_\chi \approx$ 156 MeV \cite{HotQCD:2018pds}, and
estimates of $\mu_{q,c}$, while less certain, indicate values $< $ 200
MeV \cite{Clarke:2024ugt}. Here our aim is not to tune the
absolute value of these parameters but to look at their variation with
the system size.

We will be studying the system at finite temperature $T$ and quark
number chemical potential $\muq$. $\muq$ is introduced by adding $\muq
\psi^\dagger \psi$ to \eqn{NJLlag}. To get the thermodynamic
potential, the integral over $p_0$ in \eqn{veff} has to be turned into
a fermionic Matsubara sum. For finite volume, the integral over
spatial momenta, $\dfrac{d^3p}{(2 \pi)^3}$, will be replaced by a sum
over the momentum modes allowed by the boundary condition. For a study
of the effect of finite volume, imposing a proper boundary condition
is important. As we discussed in \scn{intro}, the exact boundary
condition for the medium created in RHIC or LHC is complicated; because it is time-dependent and also changes from event to event. Since we
are only interested in getting an idea about how different the finite
volume system can be from an infinite volume system, we will take a
simple spherical geometry. The important physics that we want to
capture is that the deconfined QGP is only contained within this
geometry. For this, we impose the MIT boundary condition, which
implies that the normal component of the quark current becomes zero at
the boundary. More precisely, we impose the boundary condition
\cite{Jaffe:1989pn,Greiner:1995jn},
\beq
(-i{\hat r}\cdot {\gamma}) \psi(t,r,\theta,\phi)|_{r=R}=
\psi(t,r,\theta,\phi)|_{r=R},
\eeq{MIT}
where $\hat{r}$ is the unit vector normal to the spherical surface and
$\gamma_\mu$ are the Dirac matrices. \eqn{MIT} ensures the vanishing
of the normal component of the fermionic current and scalar densities
of quarks at the boundary of the spherical surface:
\beq
  J_n \equiv n_\mu \bar{\psi} \gamma^\mu \psi=0, \qquad \text{and}
  \qquad \bar{\psi}\psi=0 \ \ \ \text{at} \ \ \ r=R,
\eeq{MIT2}
where $\hat{n}$ is the unit normal four-vector to the boundary. The
allowed momentum modes for \eqn{MIT} are well-known; the relevant
equations are given in \apx{detail}. Then the expression for the mean-field thermodynamic potential $\Omega$ becomes:
\begin{widetext}
\beq
\hspace{-0.5cm} \Omega=\frac{(M-m)^2}{4\,G}-\frac{N_c\,N_f}{V} \sum_j \sum_{p_{j_\kappa}} (2j+1) \left[ \left\{-\frac{\Lambda\,e^{-\left(\frac{E}{\Lambda}\right)^2}}{\sqrt{\pi}}+E\,\text{Erfc}\left(\frac{E}{\Lambda}\right)\right\}+T\,\text{log}\left(1+e^{-\frac{(E-\muq)}{T}}\right)+T\,\text{log}\left(1+e^{-\frac{(E+\muq)}{T}}\right) \right],
\eeq{potfv}
\end{widetext}
where $V=\dfrac{4}{3} \pi R^3$ is the volume of the system, and
$\efc(x) = 1 - \dfrac{2}{\sqrt{\pi}} \int_0^x dy \, e^{\txt -
  y^2}$. The term in the curly bracket in \eqn{potfv} is the
vacuum term and the other two terms correspond to the
medium quark and antiquark contributions, respectively. $M$ is
determined self-consistently from the minimum of $\Omega$:
$\dfrac{\partial \Omega}{\partial M} =0$. We have used the regulator
in the vacuum term only. The medium terms are regulated by temperature
and do not need an ultraviolet regulator.

Effects of finite volume on the phase diagram have been explored in
the literature in the past. Sometimes the antiperiodic or the periodic
boundary conditions have been used for this purpose
\cite{Palhares:2009tf, Almasi:2016zqf, Klein:2017shl,
  Bernhardt:2021iql, Kovacs:2023kbv}. Note that these boundary
conditions do not capture the finite volume fireball in any way: these
are boundary conditions usually imposed in theoretical studies
(e.g. numerical studies on the lattice) when the aim is to study the
infinite volume system. Besides being easy to implement, the advantage
of these boundary conditions is that they show very small deviations
from the infinite volume system, and are therefore ideal for studies
aiming for the thermodynamic limit. Our purpose here is to estimate
the size of the deviation from the infinite volume results due to the
finite volume fireball. We do not think (anti)periodic boundary
conditions are suitable for this purpose. While the boundary condition
\eqn{MIT}, \eqn{MIT2} also do not capture the whole complexity of the
finite fireball, we believe it captures the most essential aspects of
it and gives a more realistic estimation of the size of the finite
volume effects. In \apx{bc} we compare our results with those of an
antiperiodic geometry for a similar system size. We indeed find that
the volume effects are much smaller in the antiperiodic geometries.

The finite temperature phase transition is chiral symmetry restoring
in nature, and we will track the transition using $\chc$, or
equivalently, $M$. In the absence of the symmetry-breaking term $m$,
in the full quantum theory, the chiral symmetry will be restored in
finite volume at all temperatures, due to tunneling between the
classical vacua of $\lnjl$. In Monte Carlo simulations, this will show
an instability in $\chc$. A mean-field treatment will miss the physics
of tunneling, and indicate symmetry breaking at low
temperatures. This problem is not there in the presence of the mass
term corresponding to physical light quarks, where lattice simulations
show a stable result for $\chc$ at quite small volumes down to $\mpi L
\sim 1$.

\section{Chiral transition at finite volume}
\label{sec.pdg}
We explore the chiral transition line for \eqn{mfNJL}, and its shift
with volume, for the MIT boundary condition. At $\muq=0$ the chiral
symmetry restoration is a crossover. In \fgn{cc0}, we show the
constituent mass, $M$ (\eqn{gap}), which is just a constant shift of
$- \chc$ and has the same information as $\chc$. We see a rapid but
smooth decrease in $M$ around $T \approx$ 190 MeV, indicating a
crossover. At low temperatures $\ll T_\chi$, decreasing the volume to
$R=5$ fm decreases the constituent mass by $< 5 \%$ from the infinite
volume value. At high temperatures $> 200$ MeV the chiral symmetry is
restored and the chiral condensate becomes small. Therefore $M$
becomes small, approaching $m$. In small-volume systems, the crossover
region becomes somewhat smoother, but no qualitative change from the
infinite-volume case is seen in systems as small as a 5 fm
sphere. This can be further seen in the right panel of \fgn{cc0},
where the derivative $dM/dT$ is shown. To get this plot, we used a
cubic spline interpolation of the $M$ vs $T$ data of the left
plot. The plot shows a systematic but moderate reduction in the peak
height and broadening of the peak, as the volume is reduced. The peak
position is an estimate of the crossover temperature $T_\chi$, which
changes very slowly with the radius of the sphere, $R$ (a $7$ MeV
shift from infinite volume to $R=5$ fm). The chiral susceptibility
$\chim = - \frac{\txt \partial \chc}{\txt \partial m}$ also shows a
peak at the chiral symmetry restoration point. The peak position is
close to the peak of $dM/dT$, though not exactly identical at the
crossover region, and the volume dependence is similar to that seen in
\fgn{cc0}. Here we will use the peak of $dM/dT$ as our estimate for
$T_\chi$. Of course, in the case of a true phase transition (as opposed
to a crossover), all these observables will show the same transition
temperature.

\bef
  \centering
  \begin{minipage}[b]{0.45\textwidth}
    \includegraphics[width=\textwidth]{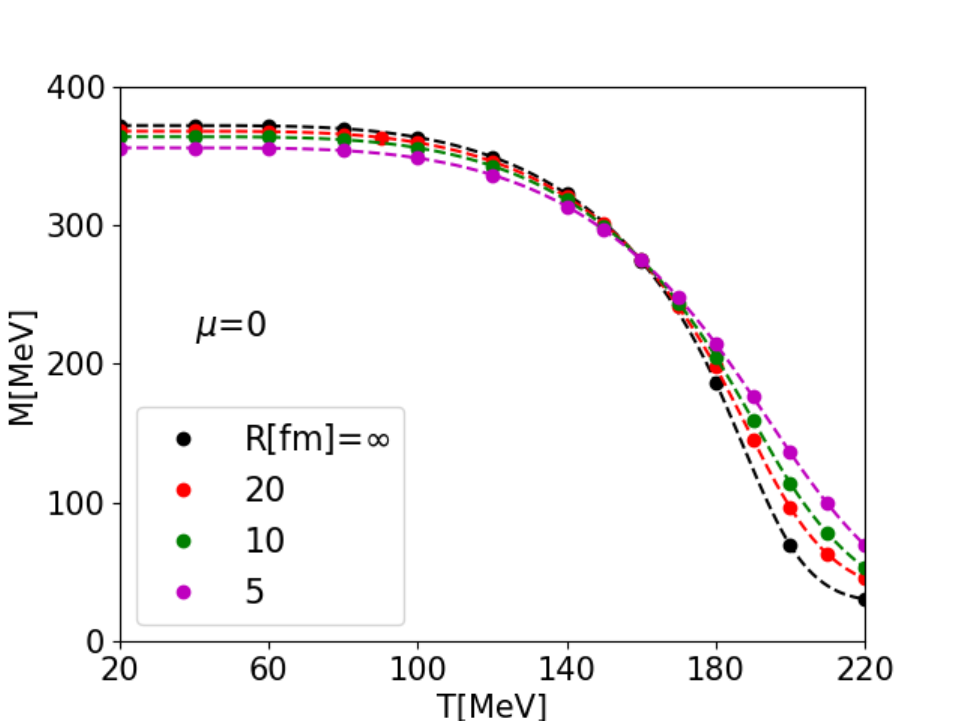}
  \end{minipage}
  \begin{minipage}[b]{0.45\textwidth}
    \includegraphics[width=\textwidth]{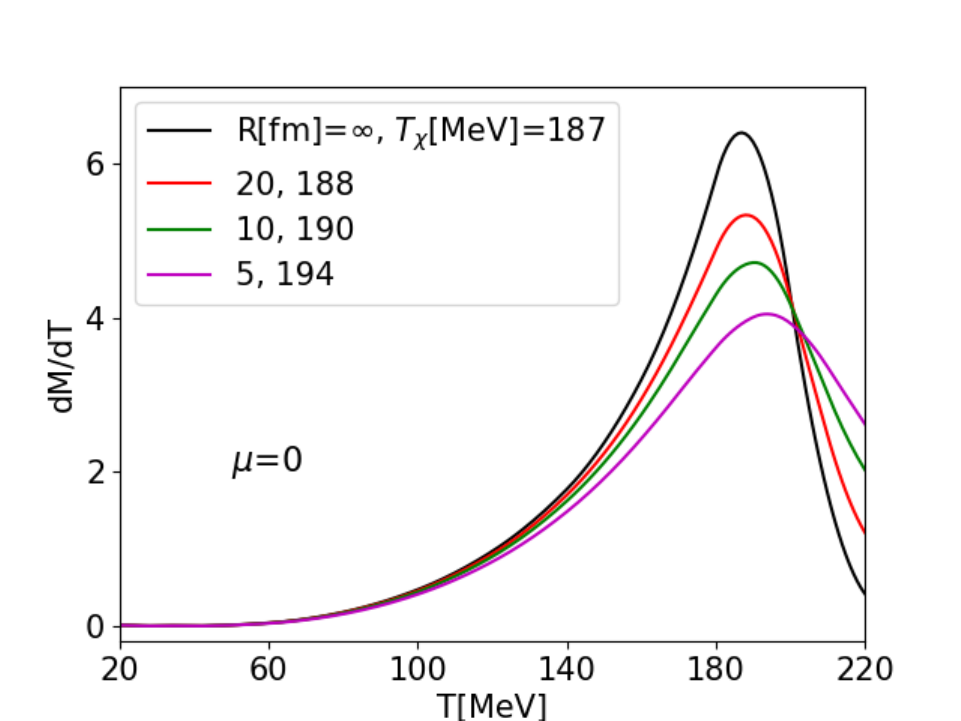}
  \end{minipage}
  \caption{\small (Left) Constituent quark mass $M$ as a function of
    temperature ($T$) at $\muq$ = 0. (Right) The derivative
    $dM/dT$. The legend shows the peak position of the $dM/dT$ curves,
    which can be taken as an estimate of $T_\chi$.}
\eef{cc0}

Now we introduce a nonzero quark chemical potential $\muq$. For small
values of $\muq$, the qualitative behavior is very similar to what is
seen for $\mu=0$. In \fgn{ccmu}, we show $M$ as a function of $T$ at
different volumes for three values of $\muq$. The temperature and
volume dependence of $M$ at $\muq$=100 MeV is very similar to that at
$\muq$=0, as a comparison of the left panel of \fgn{ccmu} with
\fgn{cc0} shows. As $\muq$ increases, the crossover shifts to smaller
values of temperature, as expected. Also, we find a stiffening of the
transition at larger volumes: the smoothing of the transition at small
volumes is enhanced. At $\muq$=350 MeV, shown in the right panel of
\fgn{ccmu}, we find a discontinuity, indicating a first-order
transition, at infinite volume. The first-order nature of the
transition can be further ascertained by looking at $\Omega$,
\fgn{pot350}. At the transition region, a two-minima structure is
seen, which is the hallmark of first-order transition, and indicates
phase coexistence. At this $\muq$ the transitions are smooth in the
finite volume system, even at $R$=20 fm, though at such a large volume
it is quite a sharp crossover. We also find here a much more enhanced
smoothing of the transition with a reduction of volume: while the
curve at $R$ = 20 fm indicates a sharp crossover, the one at $R$=5 fm
indicates a smooth behavior. Also, the crossover temperature changes
much more rapidly here with volume: the transition temperature changes
from 47.3 MeV in the infinite volume system to 72.2 MeV in the 5 fm
sphere.
\bef
\centerline{
\includegraphics[width=6cm]{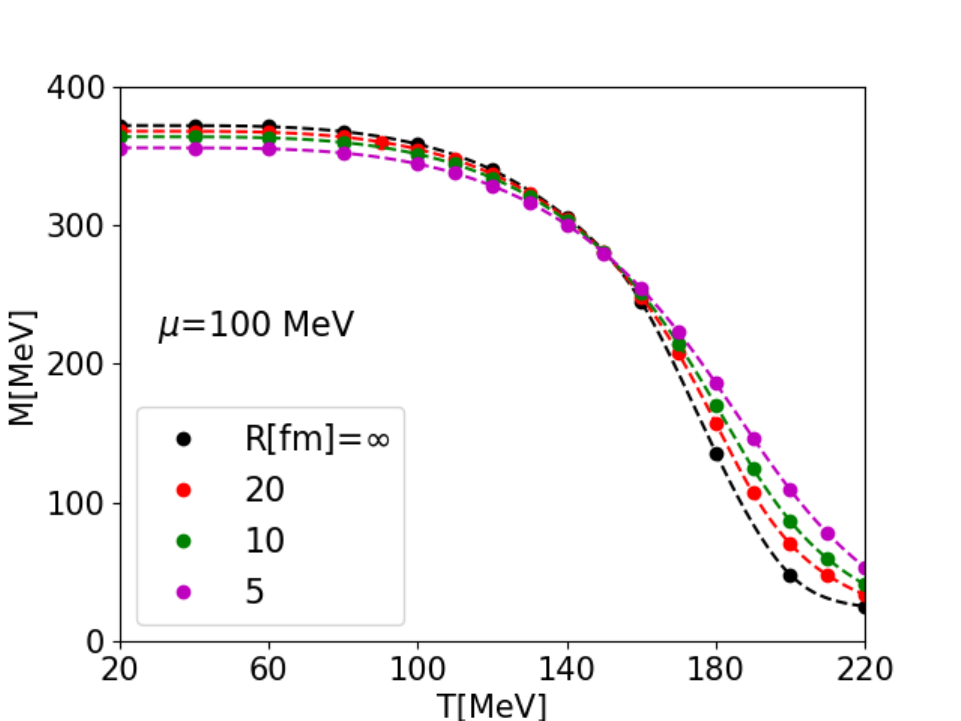}
\includegraphics[width=6cm]{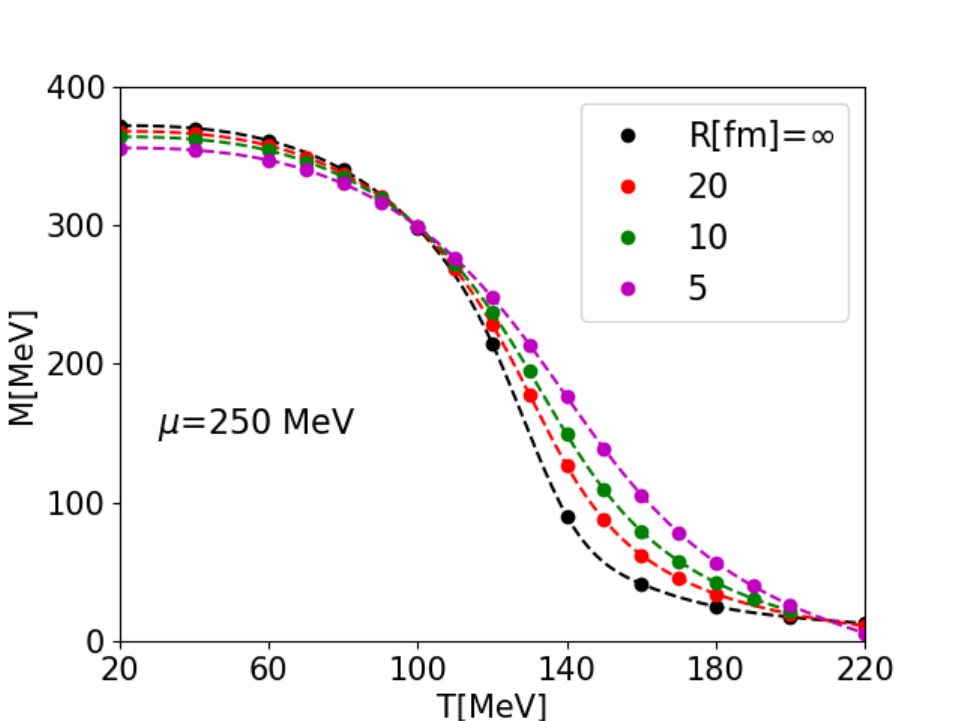}
\includegraphics[width=6cm]{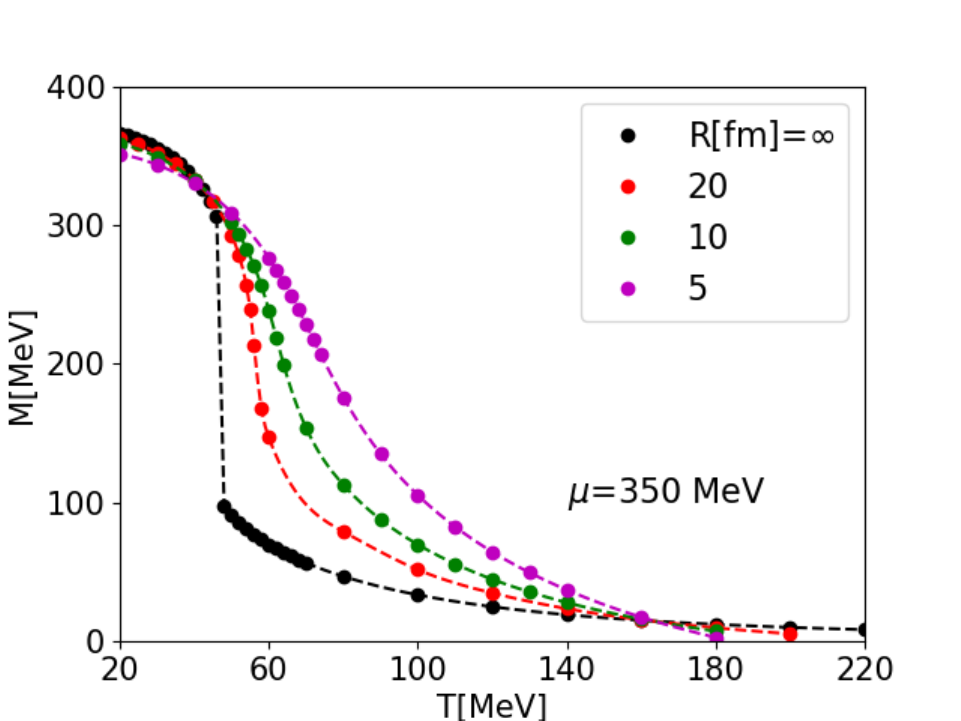}
}
\caption{\small The dependence of the $M$ vs $T$ curve with
  volume. Results at $\muq$=100 MeV (left), 250 MeV (middle) and 350
  MeV (right) are shown. A discontinuous transition is seen in the
  infinite volume system at $\muq$=350 MeV.}
\eef{ccmu}
\bef
\centerline{
\includegraphics[width=9.5cm]{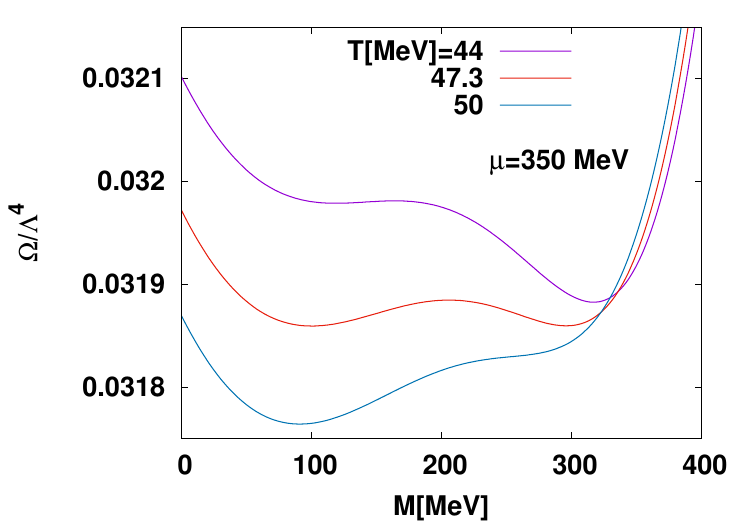}}
\caption{\small The thermodynamic potential, $\Omega/\luv^4$, in the
  transition region at $\muq$=350 MeV for an infinite volume system. A
  small (irrelevant) additive constant $\Omega=\mp 0.0001 \, \luv^4$
  has been added to the potentials at T=44 and 50 MeV, respectively,
  for better viewing. The figure indicates a first-order transition at
  $T \approx$ 47.3 MeV.}
\eef{pot350}

Before proceeding with the discussion of the investigation of the
transition line at higher $\muq$ at finite volume, let us clarify our
terminology. In our analysis, the term ``first-order transition"
corresponds to a two-well structure in the mean-field free energy, with
well-separated local minima at small and large values of $M$,
corresponding to the symmetry restored and symmetry broken phases,
respectively. At the transition point, e.g., $(T, \mu) \approx (47.3, 350)$ MeV
for the infinite volume case shown in \fgn{pot350}, $M$ shows a discontinuous jump. Such a two-minima structure is associated with phase coexistence, and
possible supercooling/superheating in a dynamical system. This is what
we have looked for, and have referred to as the signature of a
first-order transition. Similar structures at finite volume are seen at larger 
$\mu$ (see \fgn{Mhyst}).

Beyond mean-field level, we have to include fluctuations. In particular, 
in a finite volume system, the free energy barrier
between the two local minima is finite, and there is a nonzero probability for
tunneling $\sim \exp(- V \Delta \Omega/T)$ between them. This will smoothen the discontinuity in a static system. However, the tunneling rate
is exponentially suppressed in volume. In a fireball which
rapidly cools through the transition region, tunneling will be
insignificant unless the volume is very small. In cases where the
mean field free energy shows a two-minima structure and a discrete 
jump at the transition, we will
expect interesting phenomena, related to the features of a first-order
transition, at the transition point (see also Ref.\cite{Palhares:2009tf}).
Therefore, the two-well structure
of the mean-field potential, $\omf$, is of actual physical interest,
and we will loosely use terms like first-order line, phase coexistence
etc. to refer to such a structure even in a finite volume system. An
analysis of the dynamics at discontinuity in a finite volume system
has been done in Ref. \cite{Fraga:2003mu}.

\bef
\centerline{
\includegraphics[width=6cm]{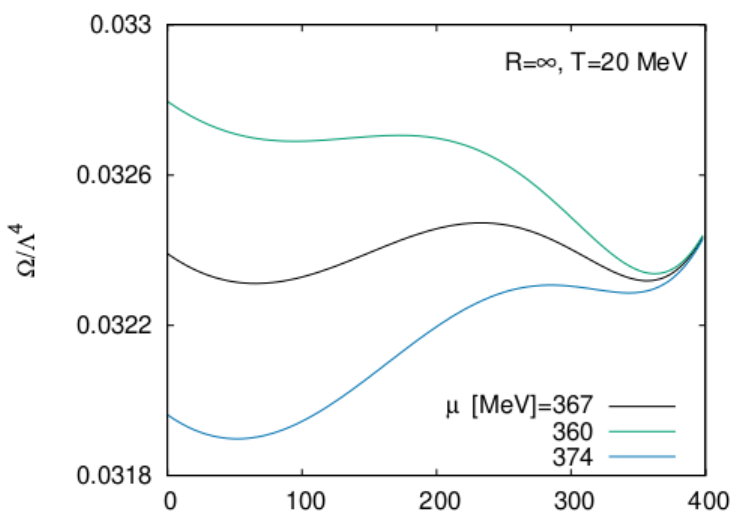}
\includegraphics[width=6cm]{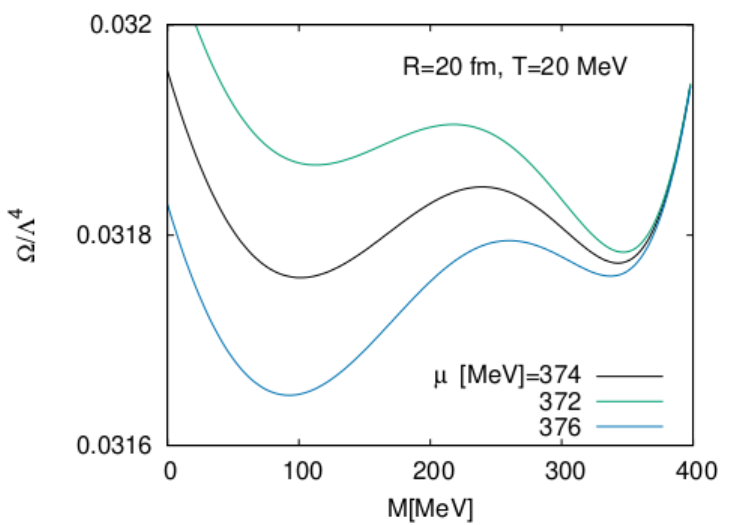}
\includegraphics[width=6cm]{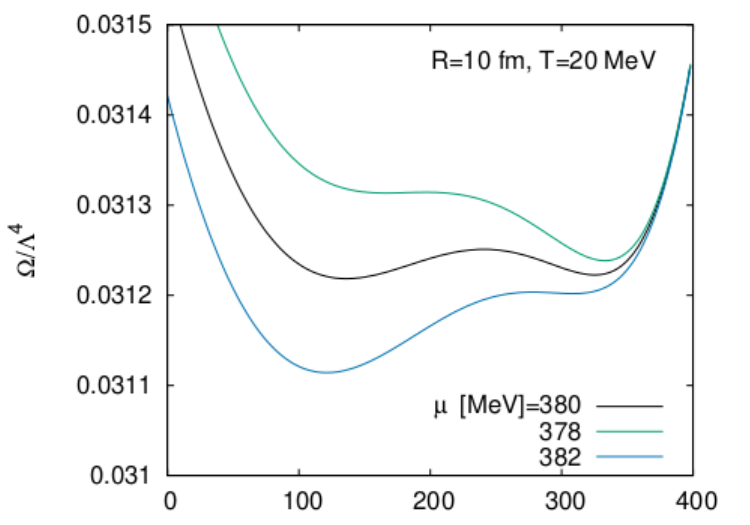}}
\caption{\small The mean-field thermodynamic potential in the
  transition region at $T$=20 MeV for infinite volume (left) and for
  spheres of radius 20 fm (middle) and 10 fm (right), showing a
  two-well structure down to a sphere of $R$=10 fm.}
\eef{Mhyst}

In \fgn{Mhyst}, $\omf$ is shown at $T$=20 MeV for different volumes.
We find a two-well structure in
$\omf$ even at $R$=10 fm. Note also the range of $\muq$ over which
this structure is seen (we loosely refer below to this region as the
coexistence region). A shift of the transition to higher $\muq$ is seen
with the reduction in volume; at the same time, a weakening of the
double-well structure (and a shrinking of the coexistence region) is
seen. At this temperature, we do not find a discontinuous transition
in the 5 fm sphere, but as we go further down to $T$=0 we find a
discontinuous transition in all our systems.

Our finding of the transition line in the $\tmq$ plane is summarized
in \fgn{pdg}. The dashed line corresponds to the smooth crossover region,
whereas the solid part of the line at small $T$ indicates the region where
the mean field effective potential behaves like a first order transition.
We find a shift of the crossover line towards larger values of $\muq$
as the system size is reduced. The shift is mild for systems with radii 
as small as 10 fm, but a more significant shift is seen for the 5 fm sphere.

A more significant volume dependence is found when we investigate the first
order transition region at low temperatures. This part is shown more clearly
in the inset. As the figure shows, the coexistence region in the mean field
potential is pushed down to lower values of $T$ as the system radius is
decreased. The dotted lines
around the first order line in the inset 
indicate the estimate of the boundary of the coexistence region. This region
shrinks quite strongly as the system size is reduced.
\bef
\centerline{\includegraphics[width=12cm]{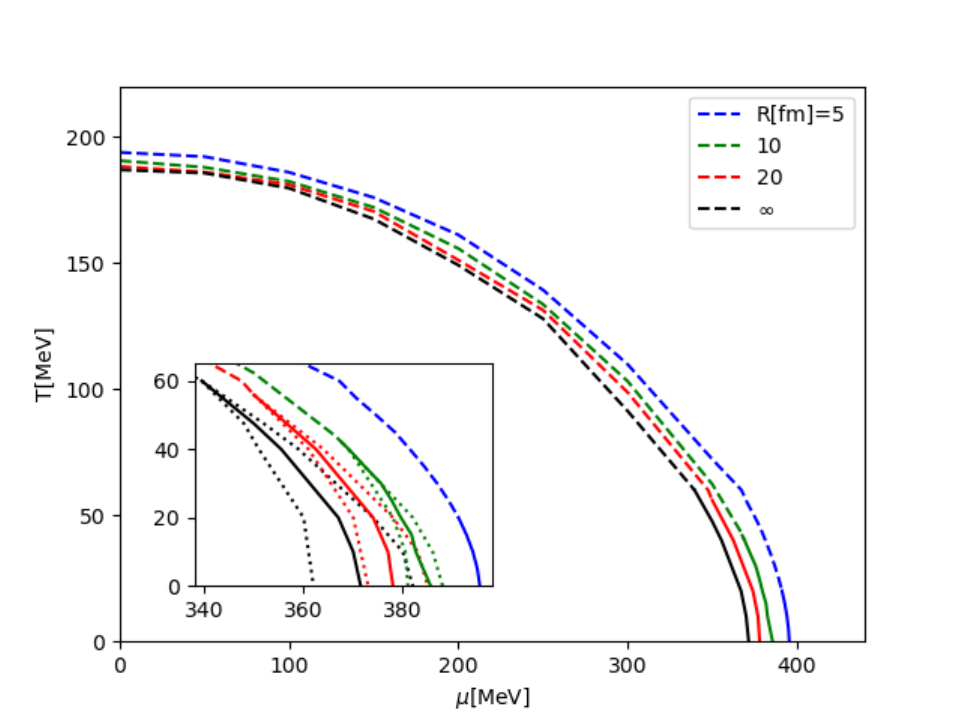}}
\caption{\small The shift of the chiral crossover line in the $(T,\muq)$
  plane with system size. In the inset the large $\muq$ part is enlarged,
  highlighting the first-order region.}
\eef{pdg}

\section{Number density and quark number susceptibility}
\label{sec.num}
The chiral condensate $\chc$ and, equivalently, the constituent mass
$M$ are good order parameters for tracking the chiral transition. They
are, however, not directly connected to an experimental
observable. The baryon number density ($\nB$) and susceptibilities of
the baryon number operator are, on the other hand, related to
experimental observables. In particular, they are of great interest in
the large $\mub$ regime, where one looks for the first order transition
line and its endpoint.

At the mean-field level the quark number density ($\nq \, = \, 3 \nB$)
is simply calculated as
\beq
n=\frac{\partial P}{\partial \mu} = 2 N_c N_f \int
\frac{d^3 p}{(2\pi)^3}\left[ \nfm(E-\muq; T) - \nfm(E+\muq; T) \right] \, ,
\eeq{nq}
where $\nfm$ is the Fermi distribution, $\nfm(x;T) = \left(e^{\tfrac{x}{T}} +
1\right)^{-1}$.
We start with an investigation of the effect of finite volume on $\nq$
in the crossover region. In the left panel of \fgn{numden}, we show
the temperature dependence of $\nq$ at $\muq$=100 MeV. At small values
of $\muq \lesssim 100$ MeV, the volume dependence of the number density is
small except in the crossover region (150 MeV $\leq \ T \ \leq$ 200
MeV). $\nq$ is small at small temperatures, and then rises and reaches
its asymptotic behavior $\sim \muq T^2$ at high temperatures. In the
crossover zone, we see some volume dependence: the rise of $\nq$ is
milder at smaller volumes. This behavior gets further amplified as we
go to higher values of $\muq$, where the crossover becomes
sharper. In the middle panel of \fgn{numden} we show the results for
$\muq$=250 MeV, still in the crossover region. Since the crossover is
now sharper, $n$ increases much more sharply across the transition
line at larger volume systems. As we enter the chirally symmetric phase, the
temperature is still not sufficiently high to show the asymptotic
$\sim \muq T^2$ behavior. As we go to even higher values of $\muq$, we
reach the first-order region, where $n$ shows a discontinuous jump
across the transition. This is seen in the infinite volume case of the
right panel of \fgn{numden} ($\muq$=350 MeV). The MIT sphere system has a smooth crossover at this value of $\muq$ for $R \le$ 10 fm, while the 20 fm sphere is close to a transition, as is also
seen in \fgn{pdg}. The behavior of $n$ above the transition is quite
different at such a large value of $\muq$, as the $\muq^4$ term
starts dominating. 
\bef \centerline{
  \includegraphics[width=6cm]{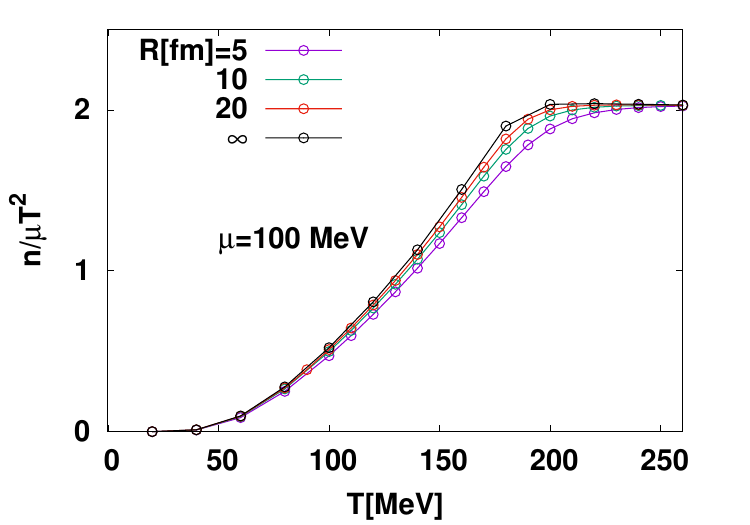}
  \includegraphics[width=6cm]{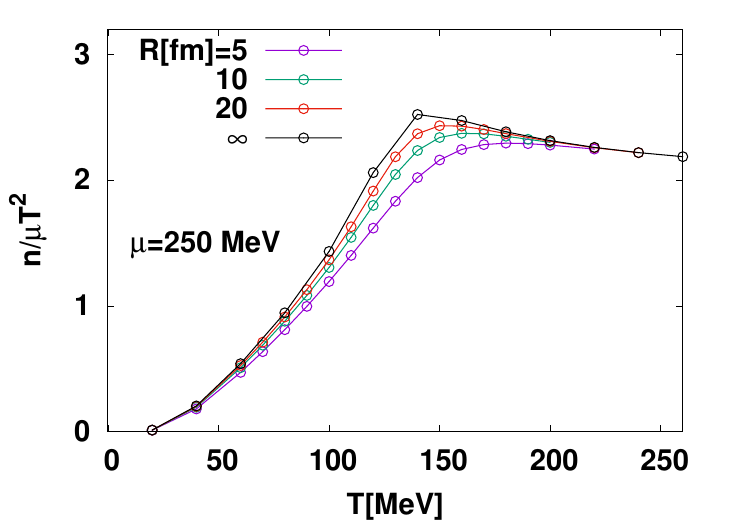}
  \includegraphics[width=6cm]{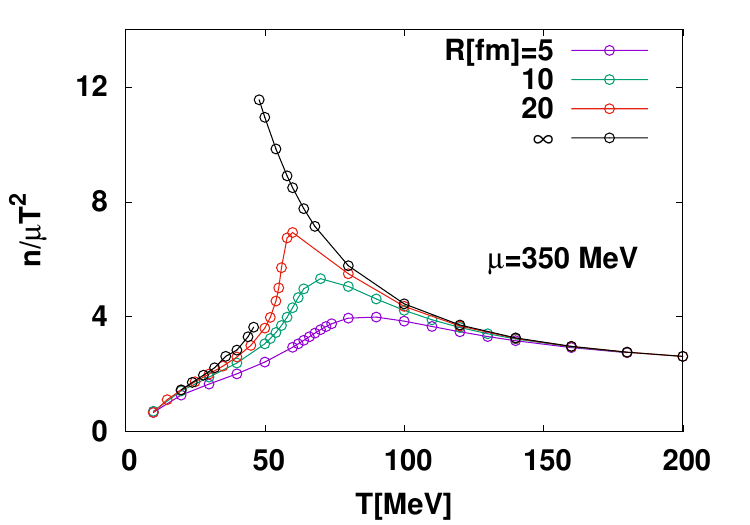}}
\caption{\small The quark number density ($\nq$) as a function of $T$
  on a system of different sizes (with MIT b.c.), at a quark chemical
  potential $\muq$ of 100 MeV (left), 250 MeV (middle) and 350 MeV
  (right).}
\eef{numden}

The quark number density, and the nonlinear quark number
susceptibilities which are higher-order derivatives of the free
energy, are very interesting observables, especially for the critical
point search. The nonlinear susceptibilities are defined as
derivatives of the pressure ($P$) with respect to the quark chemical
potential. Similar susceptibilities of other conserved quantities like
charge or strangeness can also be constructed, but here we will
concentrate on the quark number susceptibilities or equivalently,
baryon number susceptibilities. The $k$th order susceptibility
$\chi_k$ is defined as,
\beq
\chi_k=\frac{\partial^k \, P}{\partial \, \muq^k}=
\frac{\partial^{k-1} \, n}{\partial \, \muq^{k-1}} \, .
\eeq{sus}
The behavior of $\chi_k$ can be inferred from the $\muq$ dependence of
$n$, as sketched in the left panel of \fgn{chi2-RT}. At a fixed
temperature, $n$ increases with $\muq$, with a rapid rise in the transition
region. For a sufficiently high temperature $T >\tc$, the critical
temperature, one has a crossover. $\chi_2$ therefore shows a moderate peak
in this region. As the volume decreases, one expects a smoother crossover;
therefore the peak height decreases. This behavior is seen in our calculated
$\chi_2$ at different volumes; in the right panel of \fgn{chi2-RT}, we show the
results at $\muq$ = 250 MeV, which is in the crossover region. Note also that 
the peak position has shifted slightly with volume. This is because of the
shift of the crossover line, seen in \fgn{pdg}. 

\bef
\centerline{
    \includegraphics[width=8cm]{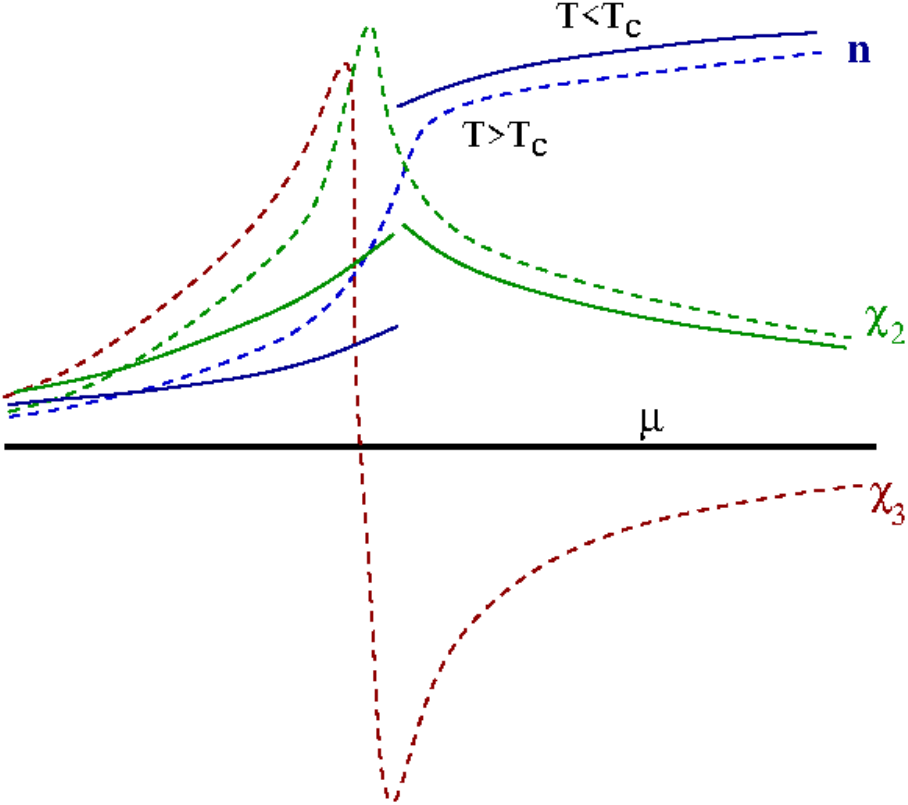} \hspace{1cm}
    \includegraphics[width=8cm]{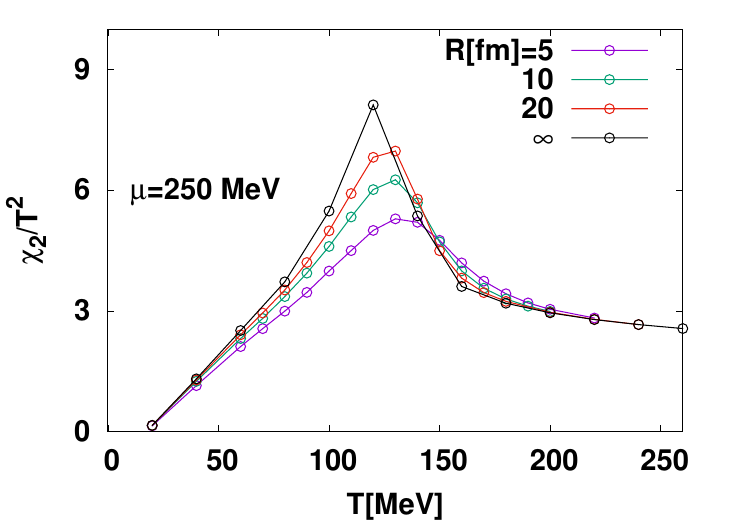}}
\caption{\small (Left) Sketch of the $\muq$ dependence of the
      number density $n$ as well as the quark number susceptibilities.
      The dashed lines show the behavior in the crossover region, at a
      temperature $T > \tc$, the critical temperature. The solid lines
      correspond to a temperature $T < \tc$, where one has a first order
      transition. (Right) The quark number susceptibility $\chi_2$ as
      a function of $T$ on a system of different sizes (with MIT b.c.),
      at a quark chemical potential $\muq$=250 MeV, which is in the
      crossover region.}
\eef{chi2-RT}

As one goes to lower temperatures, the crossover becomes steeper and
eventually one approaches the critical point, where the $\chi_2$ peak
is expected to diverge. In \fgn{chi2-3D} we show $\chi_2$ in the low
temperature region of the $\tmq$ plane, for MIT spheres with $R$=10
and 5 fm.  To obtain the curves, we calculated $n_q$ at a large number
of closely spaced values of $\muq$, fitted a cubic spline, and took its
derivative. We also checked that the values agreed with $\chi_2$
calculated directly at various points.  As we see, the $\chi_2$ peak
height increases as one approaches the critical point.  As we
discussed in \scn{pdg}, the mean field effective potential shows a
critical point also for the finite volume systems. This critical point
was seen in \fgn{pdg} to move to lower temperatures with decreasing
volume. The major volume effect seen in \fgn{chi2-3D} is related to
this shift in the critical point with volume. While the critical point
in the infinite volume system appears at $T \approx$ 60 MeV, for the
10 fm sphere one only has a smooth crossover at this temperature. The
critical point at this volume shows up at $T \approx$ 42 MeV, where
the height of the $\chi_2$ peak is maximum. At lower temperatures, one has two separate phases at
the transition: we calculate $\chi_2$ separately for the two phases,
which show moderate values. The end point of the two-state transition
happens at an even lower temperature for the 5 fm sphere, shifting the
highest $\chi_2$ peak to $T \approx$ 20 MeV.

\bef
\centerline{\includegraphics[width=12cm]{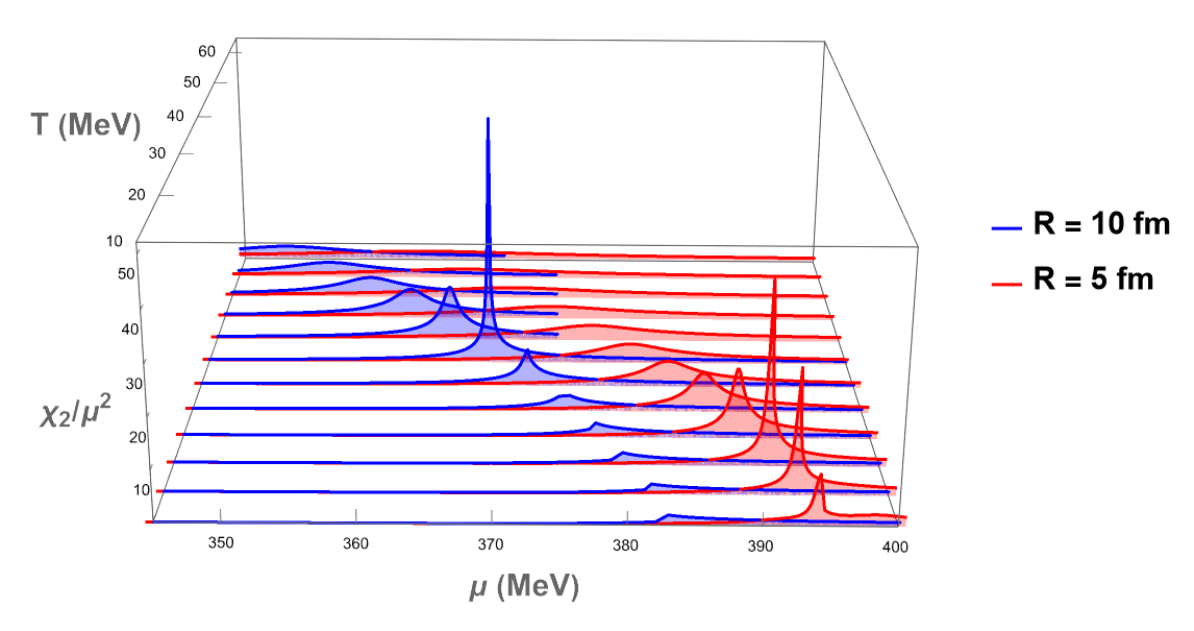}}
\caption{\small $\chi_2$ across the chiral transition line for $T
  \lesssim $ 60 MeV, which is the first order transition region for
  the infinite volume system. Shown are results for spheres of radii 5
  and 10 fm.}  \eef{chi2-3D}

It has been suggested \cite{Stephanov:2008qz, Gupta:2009mu} that the
higher-order susceptibilities are a better probe for the critical
region. In particular, $\chi_3$ \cite{Asakawa:2009aj} and $\chi_4$
\cite{Gavai:2010zn, Stephanov:2011pb} show very interesting structures
in the critical region, and these structures become smoother rapidly
with decreasing volume. The behavior of $\chi_3$ across the crossover
region has also been sketched in the left panel of \fgn{chi2-RT}.
$\chi_3$ has a peak to the lower $\muq$ side of the transition line,
then changes sign and shows a dip in the higher $\muq$ side. These
peaks and dips become sharper as one approaches the critical point.

In the left panel of \fgn{chi3}, we show the contour plot of $\chi_3$
calculated on a 10 fm sphere in the region $40 \le T \le 60$ MeV,
which is the region between the critical points on 10 fm and infinite
volume systems. The qualitative behavior of $\chi_3$ seen in the figure
is as sketched in \fgn{chi2-RT}. $\chi_3$ shows a peak as one approaches
the transition region from the low $\muq$ side, then changes sign at the
transition line and takes a large negative value just after the transition
line, before becoming small again. The peak height and the valley depth
increase as one approaches the critical region for this volume. Note that
the structure for the 10 fm sphere is quite mild at $T$=60 MeV, which
is close to the critical region for the infinite volume region. Similarly,
for the 5 fm sphere, the peak and the valley are mild at $T$=40 MeV (compare
the scales), but become steeper as one goes to lower temperatures, towards
the endpoint of the discontinuity line for this volume.

\bef
\centerline{
  \includegraphics[width=8cm]{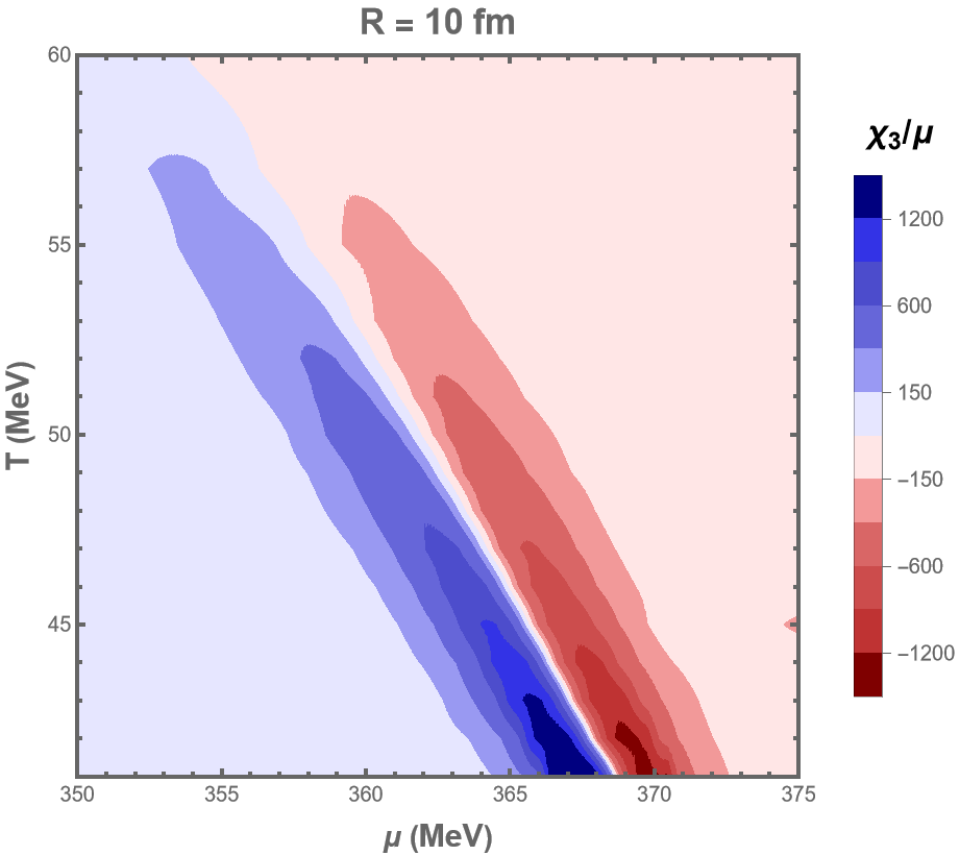} \hspace{1cm}
  \includegraphics[width=8cm]{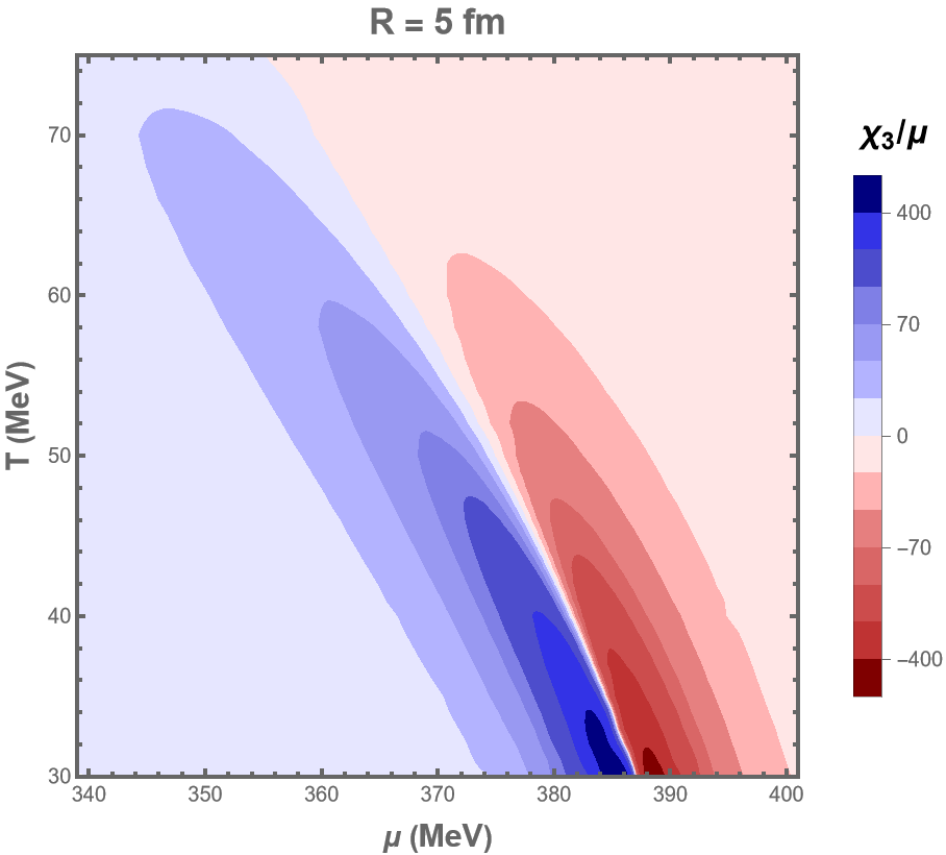}}
\caption{\small Contour plots of $\chi_3/\muq$ near the endpoint
  of the first order line, for a sphere with a radius of 10 fm (left)
  and 5 fm (right) with MIT boundary condition. We have used
  Gaussian filtering to get rid of some spurious local structures
  which are numerical artifacts.}
\eef{chi3}

In the case of a conventional boundary condition like the
antiperiodic one, the shift of the endpoint of first order line is
insignificant for a box of size 5 fm or higher (see \apx{bc}), and
the change in the magnitude of the $\chi_2$ peaks is due to the
smoother nature of the transition at smaller volumes. As \fgn{chi2-3D}
and \fgn{chi3} show, here the major effect on the peak structure is
due to the shift of the transition point. While our calculation is
only at the mean field level, we believe this effect needs to be kept
in mind when looking at the volume dependence of the baryon number
cumulants (which are closely related to the susceptibilities).

\section{Summary of results and conclusion}
\label{sec.summary}
The QCD phase diagram has been the subject of intense theoretical and
experimental investigation for a few decades. The phases at small
values of $\mub$ and the crossover line are well-understood from
non-perturbative numerical studies of QCD using lattice
discretization. The phase diagram at large values of $\mub$ has to be
studied using effective models of QCD. In particular, the Nambu
Jona-Lasinio model in its different variations has been widely used to
study the phase diagram in the $\tmu$ plane. On the experimental side,
the phase diagram is studied by creating a strongly interacting medium
at high temperatures and chemical potential in ultrarelativistic heavy
ion collisions, in particular, in the BES series of experiments at
RHIC, and other planned experiments like the NICA at JINR. To infer
the phase diagram from the experiments, it is necessary to compare
observables, in particular the number density and its
susceptibilities, with theoretical calculations.

The fireball created in the experiments has a finite volume. It is
rapidly evolving in time, and its dimensions are a few fermi before
freezeout. On the other hand, the theoretical studies are usually for
infinite volume, static systems. To infer information about the phase
diagram by combining the two, it is therefore important to estimate
the effect of the finite size and the dynamic nature of the fireball
on the observables studied. In this paper, we have studied one aspect
of it, the effect of the finite size of the system. We have studied
the 2-flavor NJL model on a sphere, with MIT boundary conditions to
mimic the fireball property of having a deconfined medium within the
finite volume. While this is an idealization of the complicated
geometry of the fireball created in a heavy ion collision, the MIT
boundary condition captures the essential feature of the finite volume
fireball that the quarks are confined within a finite
system. Therefore one can reasonably expect to learn about the effect
of the finite size using this setup.

The result of our investigation of the phase diagram can be summarized
by \fgn{pdg}, which shows the shift of the transition line with the
system volume. For the parameter sets used in our study, the NJL model
has a first-order transition line at large $\muq$, ending in a
critical point at $\left( T_c, \mu_c \right) \approx$
(60, 340) MeV. The actual location of
the transition line as well as the endpoint is sensitive to the
parameters chosen; the interest of this paper is on the sensitivity of
the line on the system volume. Here the transition line has been
tracked by the constituent quark mass $M$, or equivalently, $\chc$. As
the figure shows, in the crossover region, the shift of the transition
line is mild: even for a system with $R$=5 fm, the shift is only a few
MeV. The volume dependence of the number density and its susceptibilities
are discussed in \scn{num}.  The volume dependence of the number
density and its change across the transition line is shown in
\fgn{numden}. As expected, the change of the number density across the
crossover line becomes smoother as the volume is reduced.

The effect
of the finite volume on the transition line is more interesting in the
first-order region on the high $\muq$ side. The first-order region
(indicated by the two-minima structure of the mean-field free energy;
see Section 3) ends at a lower value of $T$ as the volume is
reduced. The number density also shows very similar behavior. The
volume dependence is particularly strong in the quark number
susceptibilities of the baryon number in the critical region, as seen
in \fgn{chi2-3D}. In particular, \fgn{chi2-3D} shows that the major
effect in the variation of the $\chi_2$ peak with volume is due to the
shift of the endpoint of the first order line. One expects a sharp
peak at the transition region near this point. As the location of the
endpoint shifts with volume, the region of large $\chi_2$ peaks also
shifts. This effect seems to be more significant than the expected
sharpening of the peak heights with volume. Similarly, $\chi_3$ shows
a very interesting behavior across the transition region, with sharp
peaks and valleys around the transition line. These structures are
seen in \fgn{chi3} to show an even stronger shift with volume, due to
the shift in the location of the critical point. One would expect even
larger effects for the higher-order susceptibilities. These results
will be relevant for the experimental searches for the critical
region, where the critical region is looked for in a small fireball.

\section{Acknowledgement}
A.S. thank Prof. M. Chernodub for a helpful discussion. The
computing resources of the Department of Theoretical Physics, TIFR,
were used for this work. We would like to thank Ajay Salve and Kapil
Ghadiali for their technical assistance. S.D. acknowledges the support
of the Department of Atomic Energy, Government of India, under Project
Identification No. RTI 4002.

\appendix
\section{Calculational details}
\label{sec.detail}
In this section, we give some formulae used in the calculations for
obtaining results of \scn{pdg} and \scn{num}, and some relevant
details. 

The chiral condensate causes the quarks to have an effective mass $M$,
and at the mean-field level, the theory can be treated like a Gaussian
theory with quarks of mass $M$. $M$ is calculated self-consistently
using the following equations,
\beq
M=m-2\,G \, \chc, \qquad \qquad \chc = - \int \mathbb{D}p \,
\text{Tr}[S(p,M)],
\eeq{gap}
where $\mathbb{D}p$ is the regularized four-momentum integral and
$S(p,M)$ is the free quark propagator matrix.

We work with the proper time regularization, in which the momentum
regulator is imposed as,
\beq
\frac{1}{p^2+M^2} \longrightarrow \int_{\tuv=1/\luv^2}^\infty \, 
d\tau \, e^{- \tau \left(p^2+M^2\right)} \, ,
\eeq{pt}
and we get the chiral condensate,
\beq
\chc \; = \; - 4 \nc \nf M \, \int \frac{d^4p}{(2 \pi)^4} 
\, \int_{\tuv}^\infty d\tau \; e^{-\tau (p^2+M^2)} \; = \; - \frac{\nf \,
 \nc \, M}{4 \pi^2} \int_{\tuv}^\infty \frac{d\tau}{\tau^2} \, e^{-M^2 \tau} \, .
\eeq{chcpt}
We can easily isolate the terms quadratic and logarithmic in $\luv$:
\beq
\chc \ = \ - \frac{\nf \, \nc \, M^3}{4 \pi^2} \left( \frac{\luv^2}{M^2} \, 
e^{- M^2/\luv^2} \; - \; \Gamma\left(0, \, \frac{M^2}{\luv^2} \right)
\right), \qquad
\Gamma(0,x) \, = \, - \gamma \, - \, \log x \, - \, \sum_{k=1}^\infty
\frac{(-x)^k}{k \, \cdot \, k!} \, ,
\eeq{chcptlo}
where $\Gamma(0,x)$ is the incomplete Gamma function, and $\gamma
\approx 0.577$ is the Euler's constant  \cite{Gradshteyn:1943cpj}.

$\lnj$ is a non-renormalizable theory, and the results depend on
$\luv$. A comparison of various regularization schemes can be found in
Ref.\,\cite{Kohyama:2015hix}. In particular, the result for $\chc$ in
dimensional regularization with $\overline{MS}$ scheme is,
\beq
\chc_{\scriptscriptstyle \overline{MS}} = -\frac{\nf \, \nc \, M^3}{4 \pi^2} 
\left(-1 \, - \, \log \frac{\overline{\mu}^2}{M^2}\right),
\eeq{chcms}
where $\overline{\mu}$ is the $\overline{MS}$ scale, and we have done the
standard $\overline{MS}$ subtraction $\dfrac{1}{\epsilon} - \gamma +
\log (4 \pi)$ from the right hand side of \eqn{chcms}. If we were to
do the standard renormalization prescription of taking $\luv \to
\infty$ in \eqn{chcptlo}, then matching to \eqn{chcms} would require a
subtraction of
\[ - \frac{\nf \, \nc \, M}{4 \pi^2} \left[ \luv^2 \, + \, M^2 \left(
  \log \frac{\overline{\mu}^2}{\luv^2} \, + \gamma \right) \right]  \]
at this order. Of course, we keep $\luv$ finite, and the results
depend on the details of the regularization.

At finite temperature, the integral over $p_0$ gets replaced by a sum
over the Matsubara modes $(2 n+1) \pi T$, leading to the mean-field
result for the thermodynamic potential:
\begin{widetext}
\beq
\hspace{-1cm} \Omega=\frac{(M-m)^2}{4\,G}-2 \, N_c\,N_f \int
\dfrac{d^3p}{(2 \pi)^3} \left[ \left\{-\frac{\Lambda\,
    e^{-\left(\frac{E}{\Lambda}\right)^2}}{\sqrt{\pi}}+E\,\text{Erfc}
  \left(\frac{E}{\Lambda}\right)\right\}+T\,\text{log}
  \left(1+e^{-\frac{(E-\muq)}{T}}\right)+T\,\text{log}
  \left(1+e^{-\frac{(E+\muq)}{T}}\right) \right].
\eeq{pot}
\end{widetext}
The chiral condensate becomes:
\beq
\chc(T, \muq) \, = \, - \frac{\nf \, \nc \, M}{4 \pi^2} \left[
  \int_{\tuv}^\infty \frac{d\tau}{\tau^2} \, e^{-M^2 \tau} \; - \;
  \frac{1}{E_p} \left( \nfm(E_p-\mu; T) \, +  \,
  \nfm(E_p+\mu; T) \right) \right],
\eeq{chctmu}
where $E_p=\sqrt{p^2+M^2}$. Note that we have only regularized the
vacuum term, as the terms with the Fermi distributions are already
regularized by the temperature scale.

The momentum modes for the MIT boundary condition on a sphere can be
obtained by solving the Dirac equation for fermions of mass $M$ and
imposing the MIT boundary condition \eqn{MIT} on the
eigenfunction. The allowed momentum modes satisfy the condition
\cite{Greiner:1995jn},
\beq
j_{l_\kappa}(pR) = - \, \text{sign}(\kappa)\left(\frac{p}{E+M}\right)
j_{l_{\bar \kappa}}(pR).
\eeq{MITcond}
Here, $j_{l}$ is $l^{th}$ order spherical Bessel function, and
\begin{equation}
  l_\kappa =
  \begin{cases} -\kappa-1 & \quad \text{for} \quad \kappa<0 \\
    \kappa & \quad \text{for} \quad \kappa>0 \end{cases}
\end{equation}

\begin{equation}
  l_{\bar \kappa} =  \begin{cases} -\kappa & \qquad \text{for}
    \quad \kappa<0 \\
    \kappa - 1 & \qquad \text{for} \quad \kappa>0.\end{cases}
\end{equation}
The energy eigenvalue for such a system is,
\begin{equation}
E\equiv E_{\kappa,i}=\sqrt{q_{\kappa,i}^2+M^2},
\end{equation}
where $q_{\kappa,i}\equiv pR$ is obtained by solving
\eqn{MITcond}. Here, $i$ is the radial excitation quantum number
($i=1,2,3,...$) and $\kappa=\pm(j+1/2)$, where $j=1/2, 3/2, ...$ is the total angular momentum quantum number. Replacing the three-momentum integral in \eqn{pot} by a sum over these modes leads to \eqn{potfv}.

\section{Comparison with other boundary conditions}
\label{sec.bc}
The QCD phase diagram at finite volume has been studied in the
literature, using the NJL model \cite{Wang:2018qyq, Zhang:2019gva} and
other effective models like the PNJL \cite{Bhattacharyya:2012rp,
  Bhattacharyya:2014uxa}, the linear sigma and the Polyakov quark-meson model
\cite{Palhares:2009tf, Magdy:2019frj, Pal:2023aai}, etc., as well as using the
Dyson-Schwinger equation \cite{Bernhardt:2021iql} and the functional
renormalization group \cite{Almasi:2016zqf}. As we have mentioned in
\scn{methods}, most of the studies have used the periodic (PBC) or
the antiperiodic (APBC) boundary condition. For a cubic box of volume
$L^3$, they correspond to momentum sum with spatial momenta $\dfrac{2
  n \pi}{L}$ and $\dfrac{(2 n + 1) \pi}{L}$, respectively. Also in
some works \cite{Bhattacharyya:2012rp, Bhattacharyya:2014uxa,
  Magdy:2019frj}, an infrared cutoff $\dfrac{\pi}{L}$ has been used in
the momentum integral to mimic a system of dimensions $L$, but the
geometry has not been specified. As we have mentioned in
\scn{methods}, we do not think such boundary conditions mimic the
finite size of the fireball. The MIT boundary condition
used here captures the essential physics, though in a very simplified form,
of the finite-size fireball.

To see how much the results with the MIT boundary condition differ
from the other boundary conditions, we have compared the size of the
finite volume effects with the different boundary conditions, by
looking at the constituent mass $M$. In the left panel of \fgn{AllBC},
we have compared
the results for the temperature dependence of $M$ at $\muq=0$
for APBC and the IR cutoff for a box of dimension $L$=5 fm, with those
obtained for a sphere with $R$=5 fm with MIT boundary condition.

As anticipated in \scn{methods}, the finite volume effect is
significantly larger with the MIT boundary than what is
seen with the other boundary
conditions. The APBC and the IR momentum cutoff show very little
deviation from the infinite volume result even at $L$=5 fm. The PBC
is expected to show even less finite volume effect than the APBC, as
has been seen in lattice studies. In the chiral symmetry broken phase,
the finite volume effect with the MIT boundary condition on a sphere
is clearly more than that with APBC and IR cutoff, despite the volume
of the sphere being about four times larger than those with a cubic
geometry. In the high-temperature symmetry restored phase the boundary
effect is substantially smaller for all the boundary
conditions.

\bef
\centerline{
  \includegraphics[width=8cm]{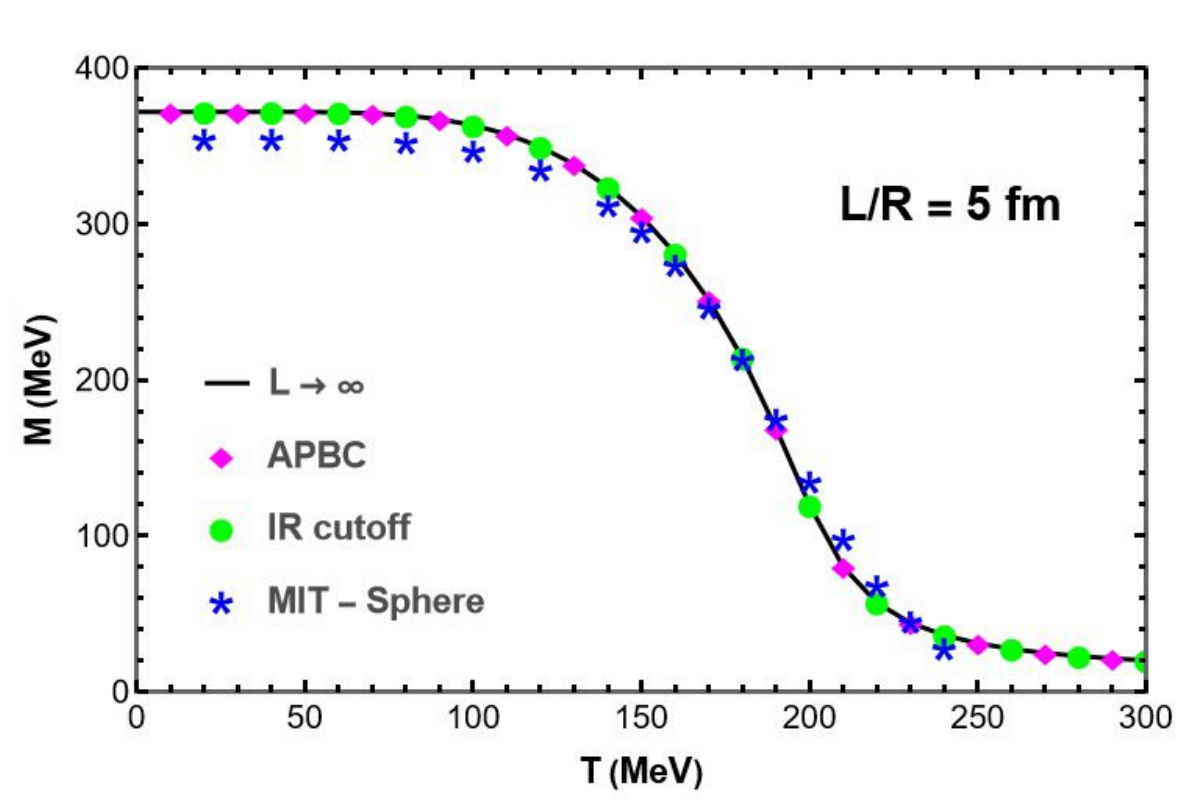} \hspace{0.5cm}
\includegraphics[width=7.75cm]{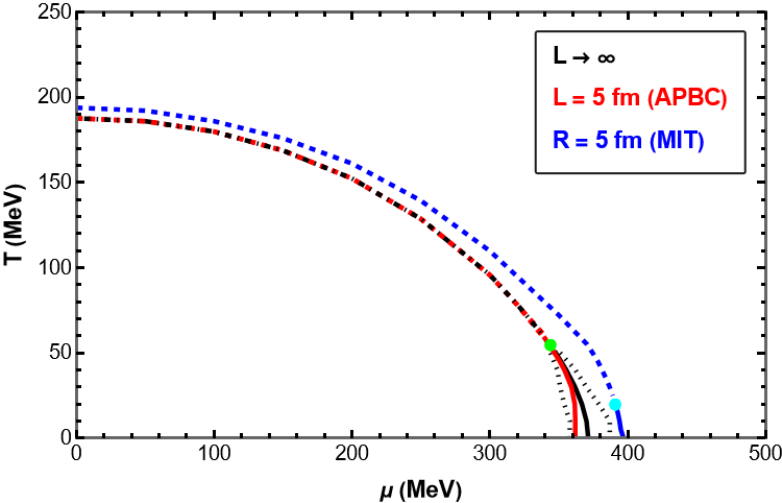}}
\caption{(Left) Constituent quark mass ($M$) as a function of
  temperature ($T$) at $\muq=0$ for different boundary conditions.
  (Right) The transition/crossover line in the $\tmq$ plane for a
  small system with MIT and antiperiodic boundary conditions, compared with
  the infinite volume line. In both these figures the MIT boundary condition
  is imposed on a sphere with radius $R$ =5 fm, while the other boundary
  conditions use a cubic box with side $L$= 5 fm.}
\eef{AllBC}

In the right panel of the figure, we have compared the phase transition line
in the $\tmq$ plane for an infinite volume system, along with those obtained
for a finite volume box of dimension 5 fm with APBC and a sphere of radius 5
fm with MIT boundary condition. The interpretation of the line is similar to
that used in \scn{pdg} and in \fgn{pdg}: it is the line obtained from the $M$
dependence of the mean field effective potential. The antiperiodic box shows
very little volume dependence even at 5 fm: the crossover line is very close
to the infinite volume one, and even the coexistence region in $M$ starts to
appear at a very similar value of $\tmq$ \footnote{For the linear sigma
model, a significant shift of the transition line, and shift of the endpoint
of the first order line towards lower temperatures, have been reported for
a 5 fm box with APBC and PBC \cite{Palhares:2009tf}.}.
At lower temperatures, the ``first
order line'' (in the mean field effective potential sense) shows slight
deviation from the infinite volume case, but still within the coexistence
region. On the other hand, as discussed in \scn{pdg}, the MIT boundary
condition leads to a substantial change: the crossover line is pushed to higher
values of $\muq$ at all temperatures. Also the coexistence region is shrunk
significantly at this volume. 
\fgn{AllBC} supports our intuition that the studies with
the APBC and similar boundary conditions significantly underestimate
the finite volume effect.
\bef
\centerline{\includegraphics[width=8cm]{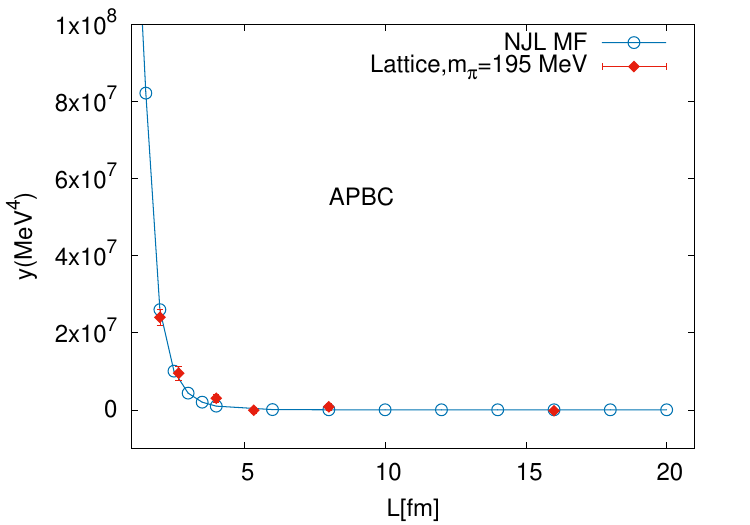}}
\caption{Comparison of $y$, \eqn{chcsub}, between lattice and MF NJL with APBC.}
\eef{lat}
While the APBC and the PBC are not suitable for studying the finite
volume fireball, they are excellent for numerical lattice studies,
where one tries to get the infinite volume results from simulations on
finite boxes. These boundary conditions are easy to implement on
lattice, and show very little volume effect, making it easy to
extrapolate to infinite volume from relatively small boxes. To check
whether the mean-field NJL model qualitatively captures the volume
dependence of the chiral condensate, we compared the results from
lattice for $\chc$ on small boxes with the NJL MF results. $\chc$ is
regularization dependent; in principle, one can use results like those
in \apx{detail} to compare the results from different regularizations.
Instead, we for simplicity
compared the renormalization group invariant quantity:
\beq
y=m\left[\langle\bar{\psi}\psi\rangle(L) -
  \langle\bar{\psi}\psi\rangle(L \rightarrow \infty)\right].
\eeq{chcsub}
The result of such a comparison is shown in \fgn{lat}. This is a
qualitative comparison where the lattice pion is somewhat heavier
($\approx$ 200 MeV). Also for the lattice, the $L \to \infty$ is
replaced by the largest lattice we used. So a precise agreement is not
expected. But as \fgn{lat} shows, the deviation from the full theory
at a very small volume is qualitatively captured in the NJL MF
result. This gives us hope that the results we obtained for the MIT
boundary condition, despite the many caveats, qualitatively capture
the trend of the finite volume effect of a static fireball.

\bibliography{reference}{}
\end{document}